\documentclass[notitlepage,superscriptaddress,showpacs,nobalancelastpage,aps,prl,twocolumn]{revtex4-1}
\usepackage[english]{babel}
\RequirePackage[T1]{fontenc}
\RequirePackage{times} 
\usepackage{amsfonts}
\usepackage{subfigure}
\usepackage{amsmath}
\usepackage{amssymb}
\usepackage{graphicx,epstopdf}
\usepackage{array}
\usepackage[hidelinks]{hyperref}
\usepackage{color}
\usepackage{my_symbols}
\usepackage{CJK}
\usepackage{bm}
\usepackage{color}
\usepackage{xcolor}
\usepackage{tikz}
\usepackage{my_symbols}
\usepackage[normalem]{ulem}

{\par}
\begin{document}
\begin{CJK*}{UTF8}{gbsn} 

\title{Antiferromagnetic domain wall motion driven by polarized spin waves}
\author{Weichao Yu (余伟超)}
\affiliation{Department of Physics and State Key Laboratory of Surface Physics, Fudan University, Shanghai 200433, China}
\author{Jin Lan (兰金)}
\email[Corresponding author:~]{jin_lan@fudan.edu.cn}
\affiliation{Department of Physics and State Key Laboratory of Surface Physics, Fudan University, Shanghai 200433, China}
\author{Jiang Xiao (萧江)}
\email[Corresponding author:~]{xiaojiang@fudan.edu.cn}
\affiliation{Department of Physics and State Key Laboratory of Surface Physics, Fudan University, Shanghai 200433, China}
\affiliation{Collaborative Innovation Center of Advanced Microstructures, Nanjing 210093, China}
\affiliation{Institute for Nanoelectronics Devices and Quantum Computing, Fudan University, Shanghai 200433, China}

\begin{abstract}
The control of magnetic domain walls is essential for the magnetic-based memory and logic applications.
 As an elementary excitation of magnetic order, spin wave is capable of moving magnetic domain walls just as the conducting electric current.
Ferromagnetic spin waves can only be right-circularly polarized.
In contrast, antiferromagnetic spin waves have full polarization degree of freedom, including both left- and right-circular polarizations, as well as all possible linear or elliptical ones.
Here we demonstrate that, due to the Dzyaloshinskii-Moriya interaction, the spin wave driven domain wall motion in antiferromagnets strongly depends on the linear polarization direction of the injected spin waves.
Steering domain wall motion by simply tuning the polarization of spin waves offers new designing principles for domain-wall based information processing devices.
\end{abstract}

\maketitle
\end{CJK*}

\emph{Introduction.}
Magnetic domain walls widely exist in all ferro-, ferri-, and antiferro-magnetic materials.
Controlling domain wall motion in a precise and convenient way is of great interest for both fundamental physics and industrial applications \cite{allwood_submicrometer_2002,allwood_magnetic_2005,parkin_magnetic_2008}.
The domain wall motion has been studied extensively, and is found to be driven by external magnetic field \cite{schryer_motion_1974, nakatani_faster_2003}, spin transfer torque \cite{berger_lowfield_1978,yamaguchi_real-space_2004,tatara_theory_2004,thiaville_micromagnetic_2005}, spin orbital torque \cite{miron_fast_2011,emori_current_2013,ryu_chiral_2013}, as well as spin waves \cite{yan_all-magnonic_2011,tveten_antiferromagnetic_2014,kim_propulsion_2014,wang_magnon-driven_2015,qaiumzadeh_controlling_2017}.
It has been shown that the antiferromagnetic domain wall can be driven more efficiently and moves much faster than its ferromagnetic counterpart \cite{yang_domain_2015,gomonay_high_2016,shiino_antiferromagnetic_2016},
which makes antiferromagnet more appealing for future magnetic-based information storage and processing techniques.

The current-driven domain wall motion via the spin transfer torque \cite{berger_lowfield_1978,yamaguchi_real-space_2004,tatara_theory_2004,thiaville_micromagnetic_2005} or spin orbital torque \cite{miron_fast_2011,emori_current_2013,ryu_chiral_2013} relies on the conduction electrons, which are not favorable for energy efficiency.
In contrast, the spin wave driven domain wall motion \cite{yan_all-magnonic_2011,tveten_antiferromagnetic_2014,kim_propulsion_2014,wang_magnon-driven_2015,qaiumzadeh_controlling_2017}
does not involve any physical motion of conduction electrons, therefore avoids the Joule heating completely.
In addition, the spin wave, as an magnetic excitation upon the static magnetization texture, is intrinsically compatible with the background magnetic texture itself \cite{lan_spin-wave_2015}. Therefore, by using static magnetic structures for memory and dynamical excitations for processing, it is possible to achieve information memory and processing simultaneously on a common magnetic system \cite{yu_logic_2017}.

With two opposite magnetic sublattices, antiferromagnetic spin waves are endowed with full polarization degrees of freedom \cite{tveten_antiferromagnetic_2014,cheng_spin_2014,cheng_antiferromagnetic_2016,lan_antiferromagnetic_2017}. Similar to photonics \cite{Goldstein_2003_polarized},
it is more natural and convenient to encode information in the polarization degree of freedom than other degrees of freedom such as amplitude or phase. Above all, the interplay between the polarized spin wave and magnetic texture in antiferromagnet would be extremely beneficial for magnonic applications.

Previously, the present authors have shown that a simple antiferromagnetic domain wall surprisingly works as a spin wave polarizer and waveplate \cite{lan_antiferromagnetic_2017}, thus is capable in manipulating spin wave polarization in full flexibility. In this Letter,  we study the equally important inverse effect, \ie the back reaction of spin wave on the domain wall, and we find that the spin wave driven antiferromagnetic domain wall motion strongly depends on the linear polarization of the injected spin waves. This polarization-dependent domain wall motion effect, in combination with its inverse process of polarizing spin waves by domain wall \cite{lan_antiferromagnetic_2017}, provides a simple scheme in exchanging binary information between the dynamical spin waves and static magnetic textures.  Our findings could lead to future logic-in-memory (or processing-in-memory) computing units solely based on magnetic systems \cite{yu_logic_2017}.

\begin{figure}[b]
\includegraphics[width=0.48\textwidth]{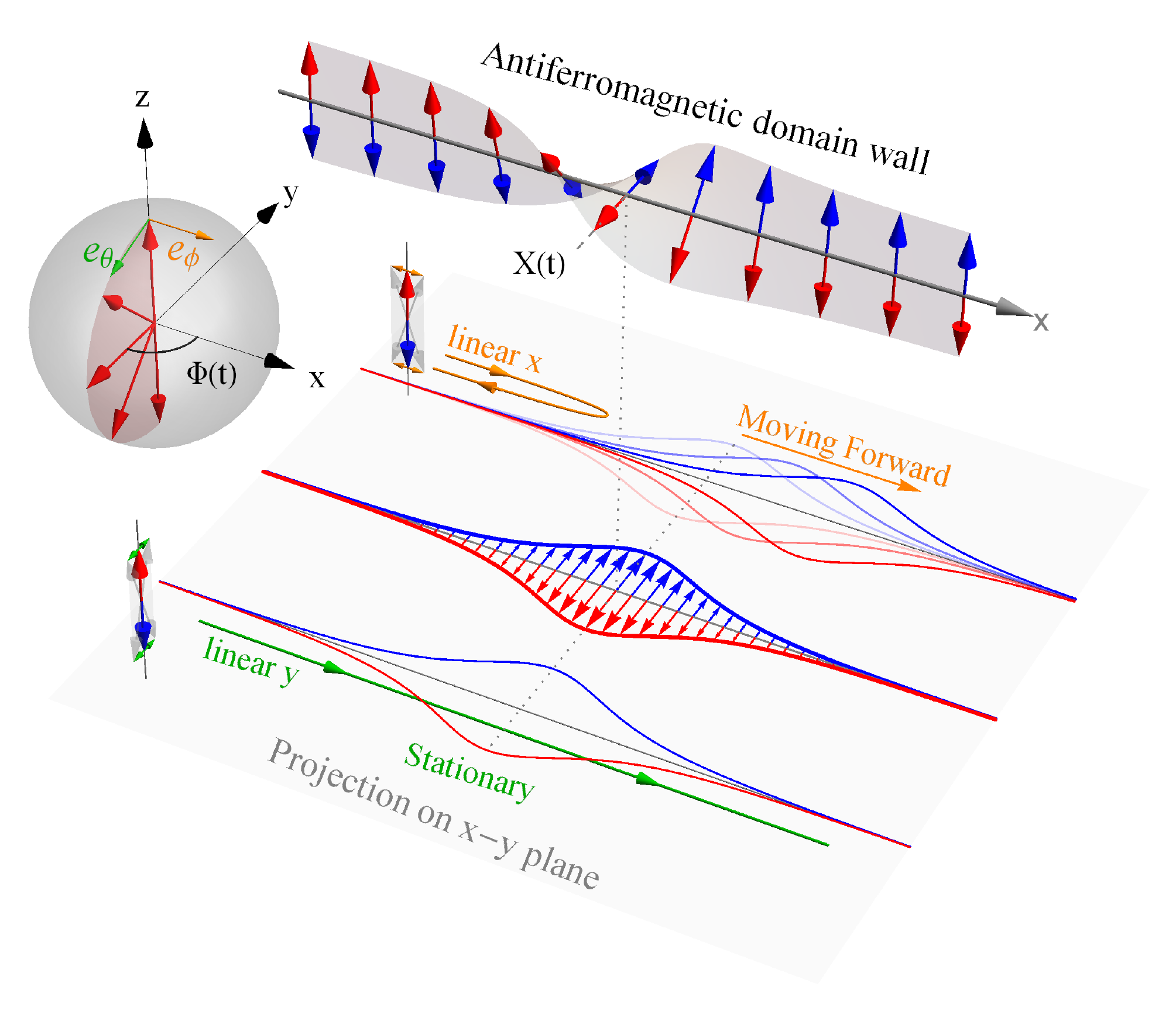}
\caption{(Color online) Schematics of the polarization-dependent domain wall motion driven by spin waves in antiferromagnet.
  An antiferromagnetic domain wall, with red (blue) arrows denoting the magnetization directions of two magnetic sublattices, is described by its central position $X(t)$ in real space and its chirality angle $\Phi(t)$ in spin space.
With DMI, the interaction between the spin waves and the domain wall is polarization-dependent: the linear-$x$ polarization spin wave (orange) is reflected by the domain wall and pushes the domain wall forward; the linear-$y$ polarization spin wave (green) transmits through the domain wall and causes nearly no motion.
}
\label{fig:fig1}
\end{figure}

\emph{Model.}
Let us consider a domain wall structure in an antiferromagnetic wire along $\hbx$ direction as shown in Fig. \ref{fig:fig1}, where the red (blue) arrows denote the magnetization directions $\mb_{1(2)}$ in the two magnetic sublattices. We define the normalized staggered N\'eel order $\bn= (\mb_1-\mb_2)/|\mb_1-\mb_2 |$ and the total magnetization $\mb=\mb_1+\mb_2$. With the constraint $\bn\cdot \mb=0$, the dyanmics of the staggered field $\bn$ is governed by \cite{tveten_staggered_2013,tveten_antiferromagnetic_2014,kim_propulsion_2014,tveten_intrinsic_2016}
\begin{equation}
\label{eqn:EOM}
J^{-1}\bn\times( \ddot{\bn} \times \bn) = \bn \times( \gamma \bH_n \times \bn) -\alpha_n \bn\times (\dbn \times \bn)
\end{equation}
and $\mb= (\dot{\bn}\times\bn)/J$ becomes a slave quantity. Here $\gamma \bH_{n}=-\partial_{\bn} U/S$ is the effective field acting on $\bn$, where $S = \mu_0 M_s\mathcal{A}/\gamma$
and $M_s$ are the angular momentum line density and the saturation magnetization of one sublattice with the cross section area $\mathcal{A}$. And $\alpha_n$ is the damping constants for the staggered field $\bn$. The free energy $U(\mb,\bn) = \half S\int\!\mathrm{d}x~ [J \mb^2+A(\nabla \bn)^2-K n_z^2+D\bn \cdot(\nabla\times \bn)]$,
where $J$ and $A$ are the inter/intra-sublattice exchange constants, $K$ is the easy-axis anisotropy (along $z$-axis), and $D$ is the strength of the Dzyaloshinskii-Moriya interaction (DMI).

We denote the static profile of the antiferromagnetic domain wall along $x$-axis by the staggered field $\bn(x)=(\sin\theta \cos\phi ,\sin\theta\sin\phi,\cos \theta)$, where $\theta(x)$ and $\phi(x)$ are the polar and azimuthal angle of $\bn$ with respect to $\hbz$.
Regardless of the presence of the DMI, an antiferromagnetic domain wall in \Figure{fig:fig1} always takes the Walker type profile like its ferromagnetic counterpart with $\theta(x)=-2\arctan[\exp((x-X)/W)]$ and $\phi = \Phi = \mathrm{const}$ \cite{yan_all-magnonic_2011,lan_spin-wave_2015,lan_antiferromagnetic_2017,tveten_antiferromagnetic_2014}, where $W =\sqrt{A/K}$ is the characteristic domain wall width and $X$ is the position of the center of the domain wall.
The magnetization directions in a domain wall along $\hbx$ trace out a plane in the Bloch sphere in spin space (see \Figure{fig:fig1}). This plane forms an angle $\Phi$ relative to $\hbx$-$\hbz$ plane.
When DMI is present, because the (bulk-type) DMI prefers neighbouring magnetizations to rotate about the $x$-axis or in the $\hby$-$\hbz$ plane, the Bloch type domain wall ($\Phi = -\pi/2$, as pictured in \Figure{fig:fig1}) has lower energy than the N\'eel type ($\Phi = 0$).
Therefore, DMI breaks the spin rotation symmetry of the domain wall in $\Phi$ with respect to $\hbz$.

\emph{Spin wave in an antiferromagnetic domain wall.} With the static domain wall profile $\bn(x)\equiv \hbe_r$, let $\td{\bn}(x,t) = \bn(x) + \delta\bn(x,t)$ and $\delta \bn(x,t)=n_\theta(x,t) \hbe_\theta +n_\phi(x,t)\hbe_\phi$ be the dynamical spin wave excitations upon the static $\bn(x)$, where $\hbe_\theta$ and $\hbe_\phi$ are the local transverse (polar and azimuthal) directions with respect to the static $\bn(x)$ as depicted in \Figure{fig:fig1}. By linearizing \Eq{eqn:EOM}, the equations of motion for the transverse component of the staggered field $n_{\theta,\phi}$ reduce to two decoupled Klein-Gordon equations \cite{tveten_antiferromagnetic_2014},
\begin{subequations}
\label{eqn:eom_sw}
\begin{align}
\mbox{in-plane}: \ddot{n}_\theta &= J \left[A \partial_x^2 - U_\ssf{K}(x)\right]n_\theta, \label{eqn:eom_sw_theta} \\
\mbox{out-of-plane}: \ddot{n}_\phi &= J \left[A \partial_x^2 - U_\ssf{K}(x) - U_\ssf{D}(x)\right]n_\phi, \label{eqn:eom_sw_phi}
\end{align}
\end{subequations}
where the effects of the inhomogeneous texture of the domain wall are transformed into two effective potentials $ U_\ssf{K}(x)$ and $ U_\ssf{D}(x)$: $ U_\ssf{K}(x)=K[1-2\mathrm{sech}^2(x/W)]$ is a potential well due to the easy-axis anisotropy along $\hbz$, and $ U_\ssf{D}(x)=(D/W)\mathrm{sech}(x/W)$ is a potential barrier due to the combined action of DMI and the inhomogeneous magnetic texture \cite{lan_antiferromagnetic_2017}.
\Eq{eqn:eom_sw} describes the behavior of the spin wave modes polarizing along the $\hbe_\theta$- and $\hbe_\phi$-direction, respectively. Since the transverse direction $\hbe_{\theta}$ ($\hbe_\phi$) is in (perpendicular to) the magnetization rotation plane as seen in \Figure{fig:fig1}, we call $n_{\theta,\phi}$ modes the in-plane and out-of-plane modes, which also coincide with the $y$- and $x$-polarized modes in the region far away from the domain wall.
Away from the domain wall, $ U_\ssf{K} = K$ and $ U_\ssf{D} = 0$, the $x$- and $y$-polarized spin wave modes are degenerated with the standard antiferromagnetic dispersion $\omega_x^2(k) = \omega_y^2(k) = JK+JAk^2$ \cite{kittel_theory_1951,keffer_theory_1952}. This dispersion indicates that antiferromagnetic spin wave can be regarded as massive relativistic particles of rest mass $m_0 = \hbar\omega_0/c_s^2$ determined by the spin wave gap $\omega_0 = \sqrt{JK}$ and the effective ``speed of light'' $c_s = \sqrt{JA}$ \cite{haldane_nonlinear_1983,kim_propulsion_2014,shiino_antiferromagnetic_2016}.

\begin{figure}[t]
\includegraphics[width=0.5\textwidth]{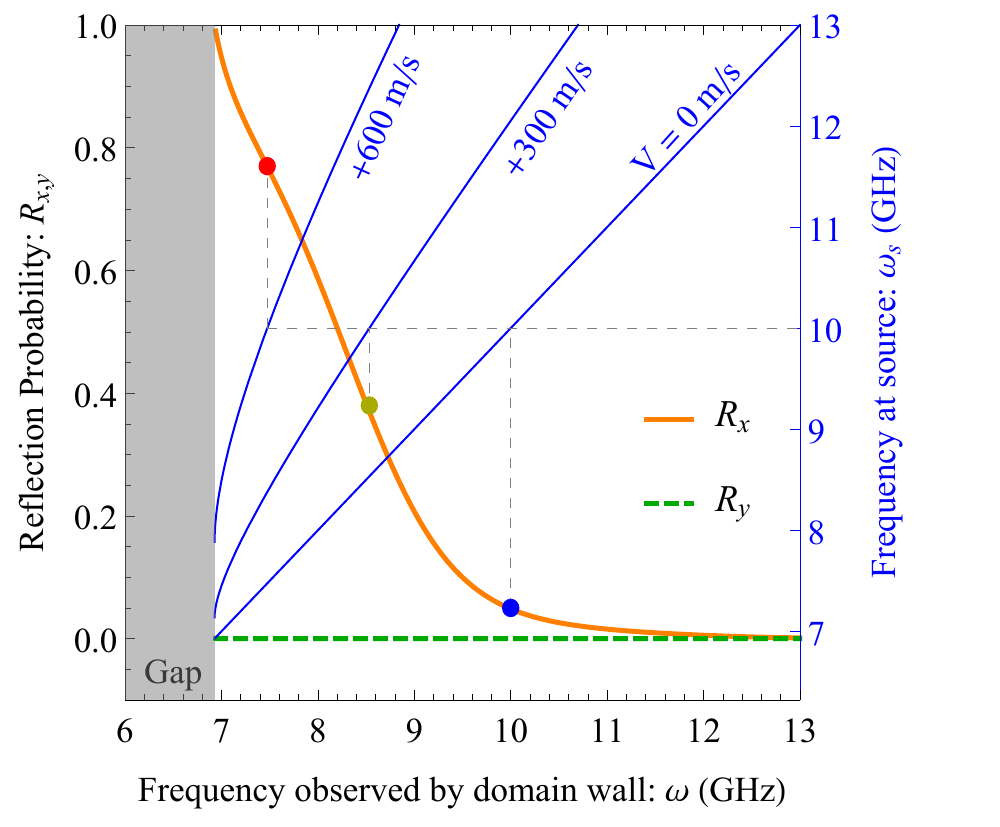}
\caption{(Color online) The frequency dependence of the spin wave reflection probability at a domain wall. The thick curves show the reflection probability for $x$-polarized (solid orange) and $y$-polarized (dashed green) spin waves.
    The thin solid (blue) curves show the Doppler shift from the frequency $\omega_s$ at the spin wave source (the right vertical axis) to the observed frequency $\omega$ (the horizontal axis) by the moving domain wall at velocity $V$. The dots are the reflection probabilities $R_x$ for spin waves of $10$ GHz as the domain wall moves with velocity $0$, $300$ m/s, and $600$ m/s.
}
\label{fig:fig2}
\end{figure}

Both effective potentials $ U_\ssf{K}(x)$ and $ U_\ssf{D}(x)$ are localized at the domain wall. $ U_\ssf{K}(x)$ in \Eq{eqn:eom_sw_theta} is the {\it reflectionless} P\"oschl-Teller type potential \cite{poschl_bemerkungen_1933}. Therefore, the (in-plane)  $y$-polarized  spin wave modes pass through the domain wall perfectly. On the other hand, the (out-of-plane)  $x$-polarized  spin wave modes are expected to experience reflections because of the additional barrier potential $ U_\ssf{D}(x)$ caused by DMI in \Eq{eqn:eom_sw_phi}. Consequently, this polarization dependent scattering at an antiferromagnetic domain wall gives rise to the polarizing effect as demonstrated in Ref. [\onlinecite{lan_antiferromagnetic_2017}]. The frequency dependence of the reflection probability  calculated from \Eq{eqn:eom_sw}  is shown in \Figure{fig:fig2}, where $R_y = 0$ identically for all frequencies and $R_x$ decreases from unity to zero as frequency increases. In practice, the reflection probability also depends on the domain wall velocity relative to the spin wave source due to the (relativistic) Doppler effect. Therefore, as the domain wall moves, the realized reflection probability relies on the observed frequency at the domain wall rather than the frequency at the spin wave source as shown in \Figure{fig:fig2}.

\begin{figure}[t]
\includegraphics[width=0.48\textwidth]{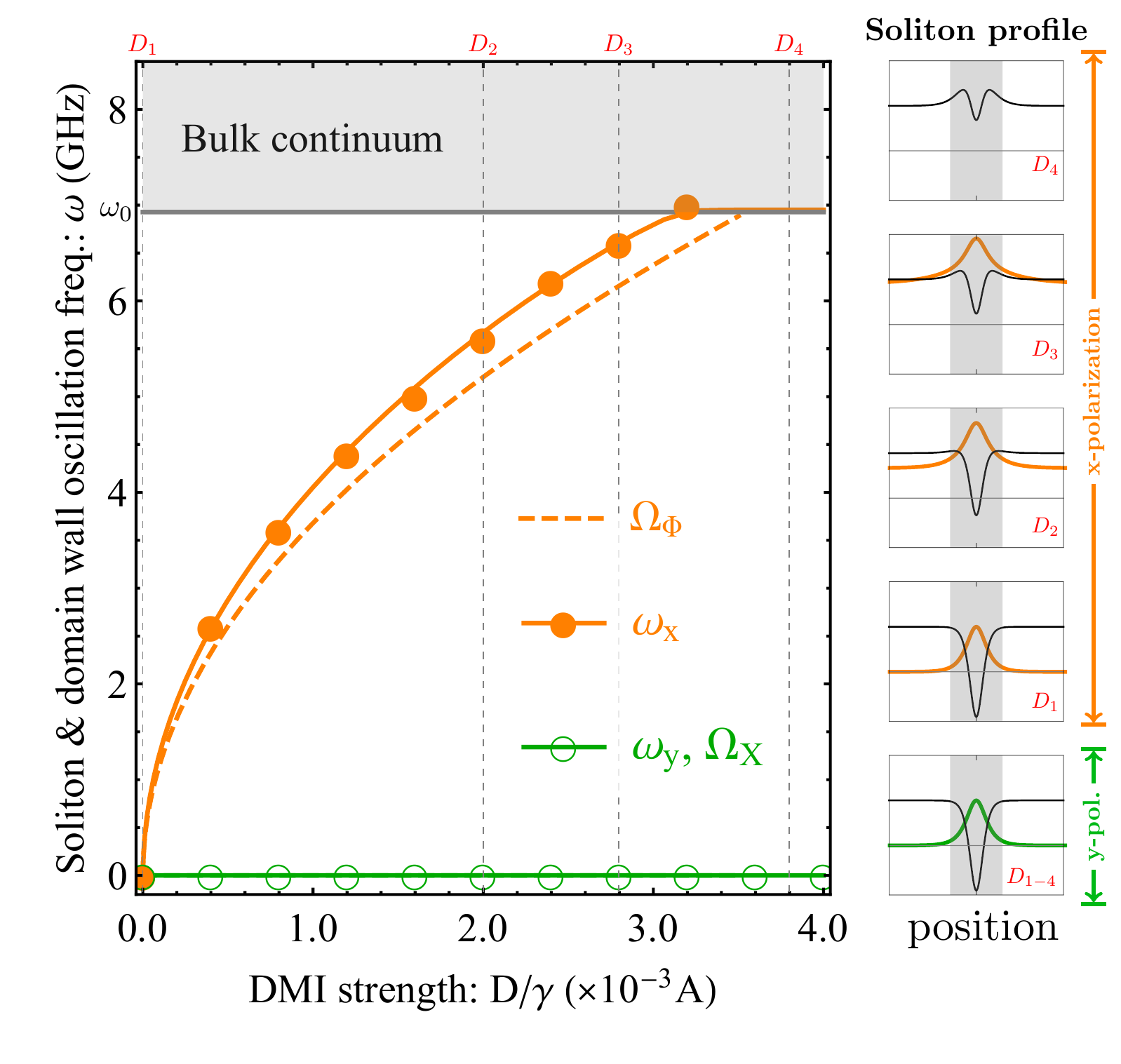}
\caption{(Color online)
 The soliton modes and domain wall oscillation modes in the presence of DMI.
Left: The soliton and domain wall oscillation frequencies as function of the DMI strength $D$.
The solid lines are soliton frequency $\omega_x$ and $\omega_y$ calculated from \Eq{eqn:eom_sw}, and the dashed lines are the oscillating domain wall frequencies $\Omega_\ssf{X} = 0$ and $\Omega_{\tiny{\Phi}} = \sqrt{2JD/\pi W}$ calculated from \Eq{eqn:Xpt}. Both the soliton and domain wall oscillation frequencies agree with the micromagnetic simulations (dots). The slight deviation of analytical $\Omega_{\tiny\Phi}$ is due to the over simplification of \Eq{eqn:Xpt}.
Right: The profiles of the potential $U_\ssf{K} + U_\ssf{D}$ (black) and the soliton (orange/green) for the $x$- and $y$-polarized soliton modes at four different values of $D$.
}
\label{fig:fig3}
\end{figure}

Apart from the scattering spin waves, \Eq{eqn:eom_sw} also hosts an in-plane and an out-of-plane soliton spin wave modes.
When DMI vanishes: $ U_\ssf{D} = 0$, both soliton modes are gapless and have zero frequency $\omega_x = \omega_y = 0$ with profile $n_{\theta,\phi}(x) \propto \mathrm{sech}(x/W)$.
When DMI is present: $ U_\ssf{D} \neq 0$, the in-plane soliton mode remain gapless at $\omega_y = 0$; but the out-of-plane soliton mode in \Eq{eqn:eom_sw_phi} becomes gapped and has a finite frequency: $\omega_x > 0$.
As DMI strength $D$ increases, the frequency $\omega_x$ approaches the bulk spin wave gap $\omega_0$, and finally merges into the bulk continuum as shown in \Figure{fig:fig3}. Such behavior can also be understood as that the out-of-plane (x-polarized) soliton mode is expelled as the potential well becomes shallower and shallower with increasing $D$, as seen in the right column of \Figure{fig:fig3}.

In fact, the two soliton modes are directly linked to the domain wall motion and rotation.
In rigid domain wall model \cite{schryer_motion_1974,tatara_theory_2004,thiaville_micromagnetic_2005,tveten_antiferromagnetic_2014,kim_propulsion_2014,wang_magnon-driven_2015,shiino_antiferromagnetic_2016,yang_domain_2015,gomonay_high_2016,qaiumzadeh_controlling_2017}, the dynamics of a domain wall can be described by the time evolution of its position $X(t)$ and rotation angle $\Phi(t)$, which represents the translation of the domain wall along $\hbx$ and the spin rotation of the domain wall about $\hbz$ respectively. In the absence of the external driving forces, by integrating over the whole space \cite{tveten_antiferromagnetic_2014,kim_propulsion_2014},
\Eq{eqn:EOM} becomes the Newtonian-like equations for $X(t)$ and $\Phi(t)$:
\begin{subequations}
\label{eqn:Xpt}
\begin{align}
M_w\ddot{X}(t) + \eta_\ssf{X} \dot{X}(t) &= 0, \label{eqn:Xt} \\
I_w\ddot{\Phi}(t) + \eta_{\tiny\Phi} \dot{\Phi}(t) &= T_{\ssf{D}}(t)\label{eqn:pt},
\end{align}
\end{subequations}
where $M_w = 2 S/(WJ)$ and $I_w = M_w W^2 = 2 SW/J$ are effective mass and moment of inertia of the antiferromagnetic domain wall, respectively. And $\eta_\ssf{X} =2 \alpha_n S/W, \eta_{\tiny\Phi} = \eta_\ssf{X}W^2$ are the viscosity for translational and rotational motion. $T_{\ssf{D}}(t) = -(4SD/\pi) \cos\Phi(t)$ is the effective pinning torque induced by DMI, whose effect is to restore the domain wall to the Bloch type ($\Phi = -\pi/2$). The decoupled dynamics for the position $X(t)$ and rotation angle $\Phi(t)$ means that domain wall translation and rotation are two independent degrees of freedom in antiferromagnet.

\Eq{eqn:Xpt} gives rise to two domain wall oscillating modes: one for the oscillation of the domain wall position $X(t)$ in real space, and the other for the oscillation of the domain wall rotation angle $\Phi(t)$ in spin space. When DMI is absent: $T_\ssf{D} = 0$, they are two Goldstone modes with vanishing frequency: $\Omega_\ssf{X} = \Omega_{\tiny{\Phi}} = 0$ and constant displacement in real and spin space: $X(t) = \const$ and $\Phi(t) = \const$ When DMI is present: the oscillating frequency for $X$ is unaffected: $\Omega_\ssf{X}=0$. But the oscillating frequency for $\Phi$ becomes finite: $\Omega_{\tiny\Phi} = \sqrt{2JD/\pi W}$ due to the pinning torque $T_\ssf{D}$, which breaks the the rotational symmetry about $\hbz$ in spin space, and always tries to restore the rotation plane back to the $\hby$-$\hbz$ plane.

The frequencies of the two oscillating modes of the antiferromagnetic domain wall coincide with that of the two soliton modes above: $\Omega_\ssf{X} \approx \omega_x$ and $\Omega_{\tiny{\Phi}} = \omega_y$. This is no accidental: the soliton picture and the rigid domain wall picture are merely two different viewpoints about the same domain wall distortion. \Figure{fig:fig3} shows that the frequencies of the soliton modes calculated from \Eq{eqn:eom_sw} and that of the oscillating modes calculated from \Eq{eqn:Xpt} indeed agree with each other, and also in agreement with the micromagnetic simulation based on the full coupled LLG equation in terms of $\mb_{1,2}$ (see Method).

\begin{figure}[t]
\includegraphics[width=8cm]{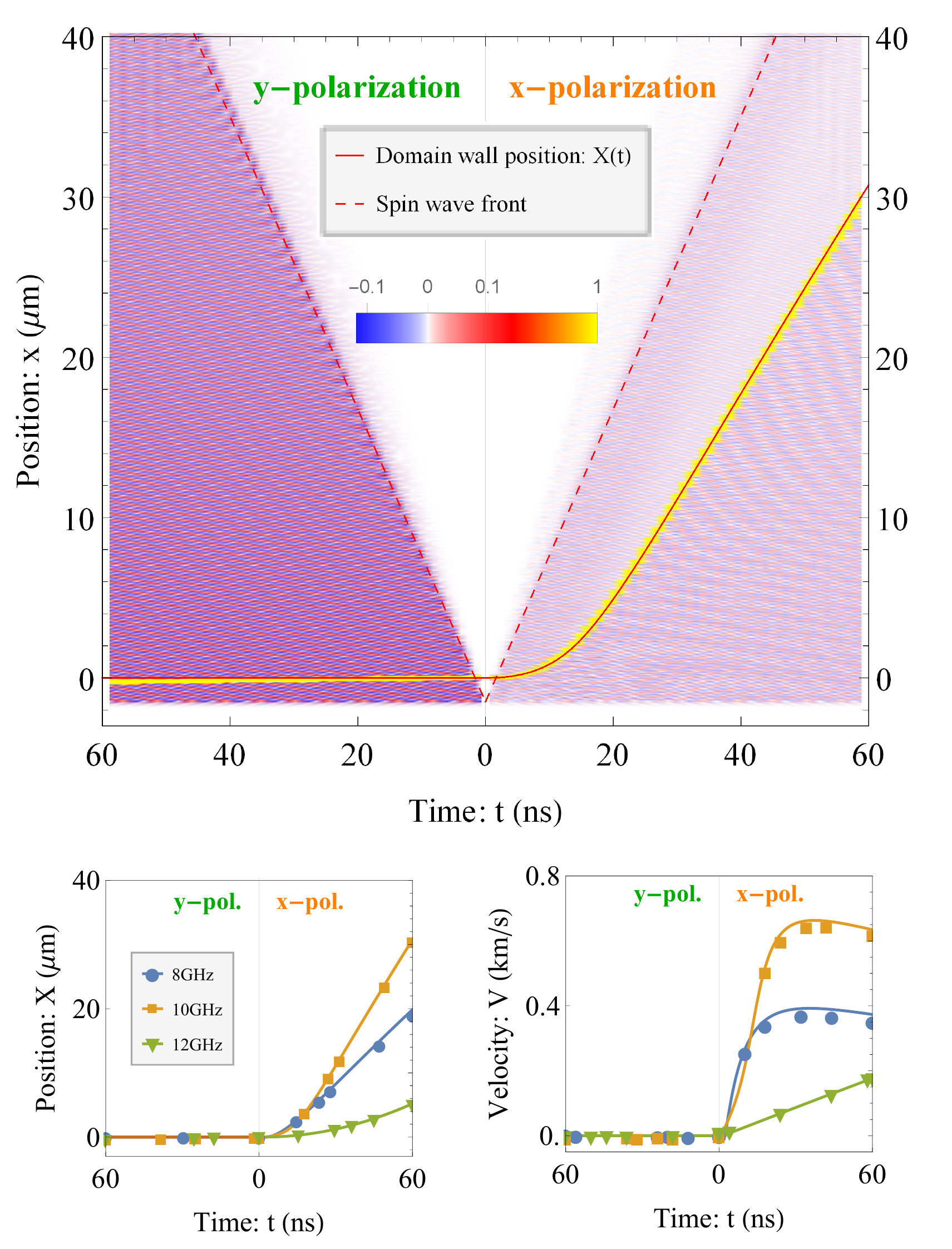}
\caption{(Color online)
The antiferromagnetic domain wall motion driven by linearly polarized spin waves.
Main panel: The magnetic profile $m^y_2$ of antiferromagnetic domain wall driven by $x$- and $y$-polarized spin wave in the full temporal and spatial range.
Spin wave is excited at the source point $x=-1.5~ \mu\mathrm{m}$ with amplitude $\rho_{x/y}=0.1$ and frequency ${\omega}=10$ GHz, and the initial location of the domain wall is $X(0)=0~ \mu\mathrm{m}$.
The time evolution of the domain wall position $X(t)$ calculated from \Eq{eqn:Xt2} is represented by the solid red curve, and the dashed red line is the spin wave front calculated using the spin wave group velocity. The domain wall width has been slightly exaggerated in the figure for a better  eye-tracking of the domain wall position.
Bottom: The domain wall position and velocity extracted from micromagnetic simulations (symbols) and calculated from \Eq{eqn:Xt2} (curves) for several spin wave frequencies $\omega =8.0, 10.0, 12.0 ~\mathrm{GHz}$.
}
\label{fig:fig4}
\end{figure}

\emph{Spin wave driven domain wall motion.}
With \Eq{eqn:Xpt}, we are ready to include the effect of driving forces. As of interests of this Letter, we consider spin wave injection, thus we need to take into account the linear momentum and the angular momentum of the spin wave transferred to the domain wall. The linear momentum is transferred via the spin wave reflection, and the resulting force pushes the domain wall forward. The angular momentum transfered gives rise to a torque driving the domain wall in rotation. In this study, we assume that the domain wall dynamics in chirality angle $\Phi$ is frozen because of the pinning torque $T_\ssf{D}$, and we only focus the dynamics in domain wall position $X$. \Eq{eqn:Xt} is modified with the spin wave force \cite{kim_propulsion_2014,tveten_antiferromagnetic_2014,qaiumzadeh_controlling_2017}
\begin{equation}
\label{eqn:Xt2}
M_w\ddot{X}(t) + \eta_\ssf{X} \dot{X}(t) = F_s(\chi,\omega,\dot{X}).
\end{equation}
The force $F_s$ exerted by the spin wave on domain wall is determined by the momentum of the reflected spin waves, which depends on the linear polarization and the frequency of the injected spin wave, as well as the domain wall velocity as shown in \Figure{fig:fig2}. Assuming that the injected spin wave is linearly polarized along $\chi = x$ or $y$ and has frequency $\omega$ and wavevector $k$, assuming that the domain wall is at rest $V \equiv \dot{X} = 0$, then \cite{kim_propulsion_2014,tveten_antiferromagnetic_2014,qaiumzadeh_controlling_2017}
\begin{equation}
\label{eqn:FT}
F_s(\chi,\omega,\dot{X}=0) = {S Ak \ov \hbar} {\rho_\chi^2(\omega)\ov 2 }R_\chi(\omega)~2\hbar k,
\end{equation}
where $\rho_\chi$ is the amplitude of the injected spin wave with polarization $\chi$, and $R_\chi$ is the reflection probability for polarization $\chi$. Because the reflection probabilities for $x$- and $y$-polarized spin waves are drastically different: $R_y = 0$ and $R_x \neq 0$, we expect that the domain wall motion should strongly depend on the spin wave polarization. \Figure{fig:fig4}(a) shows exactly that the $x$-polarized spin wave reflects at the domain wall and drives the domain wall forward, while the $y$-polarized spin wave pass through and the domain wall basically stays at rest  (also see Supplementary Movies).
 .

The force in \Eq{eqn:FT} is valid only when the domain wall does not move relative to the spin wave source, \ie $V = 0$. In order to calculate the domain wall position and velocity as function of time quantitatively, we have to take into account the relativistic Doppler effect of a moving domain wall driven by the spin wave. Therefore the frequency/wavevector used in \Eq{eqn:FT} should be replaced by their values observed by the moving domain wall: $(k,\omega)\ra (k',\omega')$ \cite{kim_propulsion_2014}:
\begin{equation}
\label{eqn:kw}
\smatrix{k'\\ \omega'/c_s}
= {1\ov \sqrt{1-\beta^2}} \smatrix{1 & -\beta \\ -\beta & 1}
\smatrix{k\\ \omega/c_s},
\end{equation}
with $\beta \equiv V/c_s$. In the mean time, the moving domain wall also becomes narrower and more massive in the lab reference frame due to the relativistic effect: $W\ra W' = W\sqrt{1-\beta^2}, M_w \ra M'_w = M_w/\sqrt{1-\beta^2}$ \cite{kim_propulsion_2014}. When including all relativistic effect of a moving domain wall in \Eq{eqn:Xt2}, we can compute the time dependence of the domain wall position $X(t)$ and velocity $V(t)$. This frequency/wavevector change is important because the reflection probability $R_\eta(\omega)$, thus the force $F_s$, strongly depends on the frequency $\omega$ seen by the domain wall (see \Figure{fig:fig2}). Therefore, the reflection probability varies in a non-trivial way as the domain wall accelerates, and the resulting domain wall velocity is a non-linear function of time.

\Figure{fig:fig4}(a) plots $m_{2}^y(x,t)$
as function of position and time obtained by micromagnetic simulations, where the cases with $x$-polarized (at right side) and $y$-polarized (at left side) spin wave injection are put side by side. The spin wave source is located at $x = -1.5~\mu$m. Much information is contained in this plot: i) the domain wall center (with $m_2^y \sim 1$) is indicated by the yellowish spot in the color plot, and the spin wave excitation is in red and blue, ii) the domain wall is transparent for $y$-polarization and opaque for $x$-polarization as expected, iii) because of ii) the domain wall moves forward with $x$-polarized spin wave, and stays at rest with $y$-polarized spin wave, iv) the wave-front travels with the spin wave group velocity as expected. The solid curve plots the domain wall position $X(t)$ calculated from \Eq{eqn:Xt2}, and the dashed line is plotted using the group velocity of the injected spin wave. Both curves agree perfectly with the micromagnetic simulations. \Figure{fig:fig4}(b) shows the comparison between results from the micromagnetic simulation and the results calculated using \Eq{eqn:Xt2} for several driving frequencies. They all agree with each other very well for both positions and velocities.

\emph{Discussions.}
The reflection of {\it linearly} polarized spin wave at the antiferromagnetic domain wall discussed in this Letter is different from the reflection of {\it circularly} polarized spin wave discussed by Tveten et al. \cite{tveten_antiferromagnetic_2014,kim_propulsion_2014,qaiumzadeh_controlling_2017} The essential difference is that, in the later case, the injected circularly polarized spin wave first drives the domain wall in precession, then the precession causes the reflection. However, in our case, the presence of DMI hinders the domain wall precession, and the $x$-polarized spin wave is reflected by the domain wall simply because DMI does not favor the $x$-polarization in the wall.

We demonstrated here the polarization dependent spin wave driven domain wall motion in antiferromagnetic wires with the bulk-type DMI. The working principle is the same for the interfacial-type DMI, for which the roles of the $x$- and $y$-polarization interchange. Furthermore, everything discussed here work in both the real antiferromagnet and the synthetic antiferromagnet consisting of two coupled ferromagnets \cite{yang_domain_2015}. For real antiferromagnet, due to the promoted exchange coupling, the spin wave group velocity and domain wall velocity can be further boosted \cite{shiino_antiferromagnetic_2016,gomonay_high_2016}.

\emph{Conclusion.}
In conclusion, we demonstrated that, due to the spin rotation symmetry broken caused by the DMI, the spin wave driven domain wall motion in antiferromagnet depends on the linear polarization direction of incident spin waves, \ie the antiferromagnetic domain wall moves with one linear polarization and stays at rest with the other orthogonal polarization. This direct connection between the polarization of the driving spin wave and the magnetic texture paves new routes in designing domain wall based magnetic logic and memory devices.

\emph{Method.}
A synthetic antiferromagnetic wire is used in micromagnetic simulations. The magnetic dynamics is described by two coupled Landau-Lifshitz-Gilbert (LLG) equations:
\begin{equation*}
\dbm_i(\br,t) = -\gamma\mb_i(\br,t)\times \bH_i^\ssf{eff}+ \alpha \mb_i(\br,t) \times \dbm_i(\br,t),
 \label{eqn:LLG}
\end{equation*}
where $i=1,2$ denote the two sublattices, $\gamma$ is the gyromagnetic ratio, $\alpha=\alpha_n/2$ is the Gilbert damping constant. Here $\gamma \bH_i^\ssf{eff}(\br,t) = \td{K} m_i^z\hbz+ \td{A} \nabla^2 \mb_i + \td{D} \nabla\times \mb_i- J \mb_{\bar{i}}$ (with $\bar{1} = 2$ and $\bar{2} = 1$) is the effective magnetic field acting locally on sublattice $\mb_i$, where $ \tilde{K}=K/2$ is the easy-axis anisotropy along $\hbz$, $\tilde{A} = A/2$ and $\tilde{D}=D/2$ are the Heisenberg and Dzyaloshinskii-Moriya exchange coupling constant within each sublattice, and $J$ is the exchange coupling constant between two sublattices.
The parameters used are as below \cite{yan_all-magnonic_2011,lan_spin-wave_2015,lan_antiferromagnetic_2017}: the easy-axis anisotropy $K=17.1\ \mathrm{GHz}$, the exchange constant $A=1.45\times10^{-5}\ \mathrm{Hz}\cdot\mathrm{m^2}$, the DMI constant $D=442\ \mathrm{Hz}\cdot\mathrm{m}$, the inter-layer coupling is chosen to $J=1.105\times10^{11}\ \mathrm{Hz}$, the saturation magnetization $M_s=1.94 \times 10^{5}\ \mathrm{A}/\mathrm{m}$, the gyromagnetic ratio $\gamma= 2.21 \times 10^{5} ~ \mathrm{Hz}/(\mathrm{A}/\mathrm{m})$, the damping $\alpha=1\times 10^{-4}$, and vacuum permeability $\mu_0= 1.26\times 10^{-6} ~\mathrm{T}\cdot\mathrm{m}/\mathrm{A} $. The dipolar interaction is neglected for this antiferromagnetic environment.
The micromagnetic simulations are performed in COMSOL Multiphysics, where the LLG equation is transformed into weak form by using the mathematical module and solved by the generalized-alpha method \cite{COMSOL}.


\begin{thebibliography}{33}%
\makeatletter
\providecommand \@ifxundefined [1]{%
 \@ifx{#1\undefined}
}%
\providecommand \@ifnum [1]{%
 \ifnum #1\expandafter \@firstoftwo
 \else \expandafter \@secondoftwo
 \fi
}%
\providecommand \@ifx [1]{%
 \ifx #1\expandafter \@firstoftwo
 \else \expandafter \@secondoftwo
 \fi
}%
\providecommand \natexlab [1]{#1}%
\providecommand \enquote  [1]{``#1''}%
\providecommand \bibnamefont  [1]{#1}%
\providecommand \bibfnamefont [1]{#1}%
\providecommand \citenamefont [1]{#1}%
\providecommand \href@noop [0]{\@secondoftwo}%
\providecommand \href [0]{\begingroup \@sanitize@url \@href}%
\providecommand \@href[1]{\@@startlink{#1}\@@href}%
\providecommand \@@href[1]{\endgroup#1\@@endlink}%
\providecommand \@sanitize@url [0]{\catcode `\\12\catcode `\$12\catcode
  `\&12\catcode `\#12\catcode `\^12\catcode `\_12\catcode `\%12\relax}%
\providecommand \@@startlink[1]{}%
\providecommand \@@endlink[0]{}%
\providecommand \url  [0]{\begingroup\@sanitize@url \@url }%
\providecommand \@url [1]{\endgroup\@href {#1}{\urlprefix }}%
\providecommand \urlprefix  [0]{URL }%
\providecommand \Eprint [0]{\href }%
\providecommand \doibase [0]{http://dx.doi.org/}%
\providecommand \selectlanguage [0]{\@gobble}%
\providecommand \bibinfo  [0]{\@secondoftwo}%
\providecommand \bibfield  [0]{\@secondoftwo}%
\providecommand \translation [1]{[#1]}%
\providecommand \BibitemOpen [0]{}%
\providecommand \bibitemStop [0]{}%
\providecommand \bibitemNoStop [0]{.\EOS\space}%
\providecommand \EOS [0]{\spacefactor3000\relax}%
\providecommand \BibitemShut  [1]{\csname bibitem#1\endcsname}%
\let\auto@bib@innerbib\@empty
\bibitem [{\citenamefont {Allwood}\ \emph {et~al.}(2002)\citenamefont
  {Allwood}, \citenamefont {Xiong}, \citenamefont {Cooke}, \citenamefont
  {Faulkner}, \citenamefont {Atkinson}, \citenamefont {Vernier},\ and\
  \citenamefont {Cowburn}}]{allwood_submicrometer_2002}%
  \BibitemOpen
  \bibfield  {author} {\bibinfo {author} {\bibfnamefont {D.~A.}\ \bibnamefont
  {Allwood}}, \bibinfo {author} {\bibfnamefont {G.}~\bibnamefont {Xiong}},
  \bibinfo {author} {\bibfnamefont {M.~D.}\ \bibnamefont {Cooke}}, \bibinfo
  {author} {\bibfnamefont {C.~C.}\ \bibnamefont {Faulkner}}, \bibinfo {author}
  {\bibfnamefont {D.}~\bibnamefont {Atkinson}}, \bibinfo {author}
  {\bibfnamefont {N.}~\bibnamefont {Vernier}}, \ and\ \bibinfo {author}
  {\bibfnamefont {R.~P.}\ \bibnamefont {Cowburn}},\ }\href {\doibase
  10.1126/science.1070595} {\bibfield  {journal} {\bibinfo  {journal}
  {Science}\ }\textbf {\bibinfo {volume} {296}},\ \bibinfo {pages} {2003}
  (\bibinfo {year} {2002})}\BibitemShut {NoStop}%
\bibitem [{\citenamefont {Allwood}\ \emph {et~al.}(2005)\citenamefont
  {Allwood}, \citenamefont {Xiong}, \citenamefont {Faulkner}, \citenamefont
  {Atkinson}, \citenamefont {Petit},\ and\ \citenamefont
  {Cowburn}}]{allwood_magnetic_2005}%
  \BibitemOpen
  \bibfield  {author} {\bibinfo {author} {\bibfnamefont {D.~A.}\ \bibnamefont
  {Allwood}}, \bibinfo {author} {\bibfnamefont {G.}~\bibnamefont {Xiong}},
  \bibinfo {author} {\bibfnamefont {C.~C.}\ \bibnamefont {Faulkner}}, \bibinfo
  {author} {\bibfnamefont {D.}~\bibnamefont {Atkinson}}, \bibinfo {author}
  {\bibfnamefont {D.}~\bibnamefont {Petit}}, \ and\ \bibinfo {author}
  {\bibfnamefont {R.~P.}\ \bibnamefont {Cowburn}},\ }\href {\doibase
  10.1126/science.1108813} {\bibfield  {journal} {\bibinfo  {journal}
  {Science}\ }\textbf {\bibinfo {volume} {309}},\ \bibinfo {pages} {1688}
  (\bibinfo {year} {2005})}\BibitemShut {NoStop}%
\bibitem [{\citenamefont {Parkin}\ \emph {et~al.}(2008)\citenamefont {Parkin},
  \citenamefont {Hayashi},\ and\ \citenamefont
  {Thomas}}]{parkin_magnetic_2008}%
  \BibitemOpen
  \bibfield  {author} {\bibinfo {author} {\bibfnamefont {S.~S.~P.}\
  \bibnamefont {Parkin}}, \bibinfo {author} {\bibfnamefont {M.}~\bibnamefont
  {Hayashi}}, \ and\ \bibinfo {author} {\bibfnamefont {L.}~\bibnamefont
  {Thomas}},\ }\href {\doibase 10.1126/science.1145799} {\bibfield  {journal}
  {\bibinfo  {journal} {Science}\ }\textbf {\bibinfo {volume} {320}},\ \bibinfo
  {pages} {190} (\bibinfo {year} {2008})}\BibitemShut {NoStop}%
\bibitem [{\citenamefont {Schryer}\ and\ \citenamefont
  {Walker}(1974)}]{schryer_motion_1974}%
  \BibitemOpen
  \bibfield  {author} {\bibinfo {author} {\bibfnamefont {N.~L.}\ \bibnamefont
  {Schryer}}\ and\ \bibinfo {author} {\bibfnamefont {L.~R.}\ \bibnamefont
  {Walker}},\ }\href {\doibase 10.1063/1.1663252} {\bibfield  {journal}
  {\bibinfo  {journal} {J. Appl. Phys.}\ }\textbf {\bibinfo {volume} {45}},\
  \bibinfo {pages} {5406} (\bibinfo {year} {1974})}\BibitemShut {NoStop}%
\bibitem [{\citenamefont {Nakatani}\ \emph {et~al.}(2003)\citenamefont
  {Nakatani}, \citenamefont {Thiaville},\ and\ \citenamefont
  {Miltat}}]{nakatani_faster_2003}%
  \BibitemOpen
  \bibfield  {author} {\bibinfo {author} {\bibfnamefont {Y.}~\bibnamefont
  {Nakatani}}, \bibinfo {author} {\bibfnamefont {A.}~\bibnamefont {Thiaville}},
  \ and\ \bibinfo {author} {\bibfnamefont {J.}~\bibnamefont {Miltat}},\ }\href
  {\doibase 10.1038/nmat931} {\bibfield  {journal} {\bibinfo  {journal} {Nat.
  Mater.}\ }\textbf {\bibinfo {volume} {2}},\ \bibinfo {pages} {521} (\bibinfo
  {year} {2003})}\BibitemShut {NoStop}%
\bibitem [{\citenamefont {Berger}(1978)}]{berger_lowfield_1978}%
  \BibitemOpen
  \bibfield  {author} {\bibinfo {author} {\bibfnamefont {L.}~\bibnamefont
  {Berger}},\ }\href {\doibase 10.1063/1.324716} {\bibfield  {journal}
  {\bibinfo  {journal} {J. Appl. Phys.}\ }\textbf {\bibinfo {volume} {49}},\
  \bibinfo {pages} {2156} (\bibinfo {year} {1978})}\BibitemShut {NoStop}%
\bibitem [{\citenamefont {Yamaguchi}\ \emph {et~al.}(2004)\citenamefont
  {Yamaguchi}, \citenamefont {Ono}, \citenamefont {Nasu}, \citenamefont
  {Miyake}, \citenamefont {Mibu},\ and\ \citenamefont
  {Shinjo}}]{yamaguchi_real-space_2004}%
  \BibitemOpen
  \bibfield  {author} {\bibinfo {author} {\bibfnamefont {A.}~\bibnamefont
  {Yamaguchi}}, \bibinfo {author} {\bibfnamefont {T.}~\bibnamefont {Ono}},
  \bibinfo {author} {\bibfnamefont {S.}~\bibnamefont {Nasu}}, \bibinfo {author}
  {\bibfnamefont {K.}~\bibnamefont {Miyake}}, \bibinfo {author} {\bibfnamefont
  {K.}~\bibnamefont {Mibu}}, \ and\ \bibinfo {author} {\bibfnamefont
  {T.}~\bibnamefont {Shinjo}},\ }\href {\doibase 10.1103/PhysRevLett.92.077205}
  {\bibfield  {journal} {\bibinfo  {journal} {Phys. Rev. Lett.}\ }\textbf
  {\bibinfo {volume} {92}},\ \bibinfo {pages} {077205} (\bibinfo {year}
  {2004})}\BibitemShut {NoStop}%
\bibitem [{\citenamefont {Tatara}\ and\ \citenamefont
  {Kohno}(2004)}]{tatara_theory_2004}%
  \BibitemOpen
  \bibfield  {author} {\bibinfo {author} {\bibfnamefont {G.}~\bibnamefont
  {Tatara}}\ and\ \bibinfo {author} {\bibfnamefont {H.}~\bibnamefont {Kohno}},\
  }\href {\doibase 10.1103/PhysRevLett.92.086601} {\bibfield  {journal}
  {\bibinfo  {journal} {Phys. Rev. Lett.}\ }\textbf {\bibinfo {volume} {92}},\
  \bibinfo {pages} {086601} (\bibinfo {year} {2004})}\BibitemShut {NoStop}%
\bibitem [{\citenamefont {Thiaville}\ \emph {et~al.}(2005)\citenamefont
  {Thiaville}, \citenamefont {Nakatani}, \citenamefont {Miltat},\ and\
  \citenamefont {Suzuki}}]{thiaville_micromagnetic_2005}%
  \BibitemOpen
  \bibfield  {author} {\bibinfo {author} {\bibfnamefont {A.}~\bibnamefont
  {Thiaville}}, \bibinfo {author} {\bibfnamefont {Y.}~\bibnamefont {Nakatani}},
  \bibinfo {author} {\bibfnamefont {J.}~\bibnamefont {Miltat}}, \ and\ \bibinfo
  {author} {\bibfnamefont {Y.}~\bibnamefont {Suzuki}},\ }\href@noop {}
  {\bibfield  {journal} {\bibinfo  {journal} {Euro. Phys. Lett.}\ }\textbf
  {\bibinfo {volume} {69}},\ \bibinfo {pages} {990} (\bibinfo {year}
  {2005})}\BibitemShut {NoStop}%
\bibitem [{\citenamefont {Miron}\ \emph {et~al.}(2011)\citenamefont {Miron},
  \citenamefont {Moore}, \citenamefont {Szambolics}, \citenamefont
  {Buda-Prejbeanu}, \citenamefont {Auffret}, \citenamefont {Rodmacq},
  \citenamefont {Pizzini}, \citenamefont {Vogel}, \citenamefont {Bonfim},
  \citenamefont {Schuhl},\ and\ \citenamefont {Gaudin}}]{miron_fast_2011}%
  \BibitemOpen
  \bibfield  {author} {\bibinfo {author} {\bibfnamefont {I.~M.}\ \bibnamefont
  {Miron}}, \bibinfo {author} {\bibfnamefont {T.}~\bibnamefont {Moore}},
  \bibinfo {author} {\bibfnamefont {H.}~\bibnamefont {Szambolics}}, \bibinfo
  {author} {\bibfnamefont {L.~D.}\ \bibnamefont {Buda-Prejbeanu}}, \bibinfo
  {author} {\bibfnamefont {S.}~\bibnamefont {Auffret}}, \bibinfo {author}
  {\bibfnamefont {B.}~\bibnamefont {Rodmacq}}, \bibinfo {author} {\bibfnamefont
  {S.}~\bibnamefont {Pizzini}}, \bibinfo {author} {\bibfnamefont
  {J.}~\bibnamefont {Vogel}}, \bibinfo {author} {\bibfnamefont
  {M.}~\bibnamefont {Bonfim}}, \bibinfo {author} {\bibfnamefont
  {A.}~\bibnamefont {Schuhl}}, \ and\ \bibinfo {author} {\bibfnamefont
  {G.}~\bibnamefont {Gaudin}},\ }\href {\doibase 10.1038/nmat3020} {\bibfield
  {journal} {\bibinfo  {journal} {Nat. Mater.}\ }\textbf {\bibinfo {volume}
  {10}},\ \bibinfo {pages} {419} (\bibinfo {year} {2011})}\BibitemShut
  {NoStop}%
\bibitem [{\citenamefont {Emori}\ \emph {et~al.}(2013)\citenamefont {Emori},
  \citenamefont {Bauer}, \citenamefont {Ahn}, \citenamefont {Martinez},\ and\
  \citenamefont {Beach}}]{emori_current_2013}%
  \BibitemOpen
  \bibfield  {author} {\bibinfo {author} {\bibfnamefont {S.}~\bibnamefont
  {Emori}}, \bibinfo {author} {\bibfnamefont {U.}~\bibnamefont {Bauer}},
  \bibinfo {author} {\bibfnamefont {S.-M.}\ \bibnamefont {Ahn}}, \bibinfo
  {author} {\bibfnamefont {E.}~\bibnamefont {Martinez}}, \ and\ \bibinfo
  {author} {\bibfnamefont {G.~S.~D.}\ \bibnamefont {Beach}},\ }\href@noop {}
  {\bibfield  {journal} {\bibinfo  {journal} {Nat. Mater.}\ }\textbf {\bibinfo
  {volume} {12}},\ \bibinfo {pages} {611} (\bibinfo {year} {2013})}\BibitemShut
  {NoStop}%
\bibitem [{\citenamefont {Ryu}\ \emph {et~al.}(2013)\citenamefont {Ryu},
  \citenamefont {Thomas}, \citenamefont {Yang},\ and\ \citenamefont
  {Parkin}}]{ryu_chiral_2013}%
  \BibitemOpen
  \bibfield  {author} {\bibinfo {author} {\bibfnamefont {K.-S.}\ \bibnamefont
  {Ryu}}, \bibinfo {author} {\bibfnamefont {L.}~\bibnamefont {Thomas}},
  \bibinfo {author} {\bibfnamefont {S.-H.}\ \bibnamefont {Yang}}, \ and\
  \bibinfo {author} {\bibfnamefont {S.}~\bibnamefont {Parkin}},\ }\href
  {\doibase 10.1038/nnano.2013.102} {\bibfield  {journal} {\bibinfo  {journal}
  {Nat. Nano.}\ }\textbf {\bibinfo {volume} {8}},\ \bibinfo {pages} {527}
  (\bibinfo {year} {2013})}\BibitemShut {NoStop}%
\bibitem [{\citenamefont {Yan}\ \emph {et~al.}(2011)\citenamefont {Yan},
  \citenamefont {Wang},\ and\ \citenamefont {Wang}}]{yan_all-magnonic_2011}%
  \BibitemOpen
  \bibfield  {author} {\bibinfo {author} {\bibfnamefont {P.}~\bibnamefont
  {Yan}}, \bibinfo {author} {\bibfnamefont {X.~S.}\ \bibnamefont {Wang}}, \
  and\ \bibinfo {author} {\bibfnamefont {X.~R.}\ \bibnamefont {Wang}},\
  }\href@noop {} {\bibfield  {journal} {\bibinfo  {journal} {Phys. Rev. Lett.}\
  }\textbf {\bibinfo {volume} {107}},\ \bibinfo {pages} {177207} (\bibinfo
  {year} {2011})}\BibitemShut {NoStop}%
\bibitem [{\citenamefont {Tveten}\ \emph {et~al.}(2014)\citenamefont {Tveten},
  \citenamefont {Qaiumzadeh},\ and\ \citenamefont
  {Brataas}}]{tveten_antiferromagnetic_2014}%
  \BibitemOpen
  \bibfield  {author} {\bibinfo {author} {\bibfnamefont {E.~G.}\ \bibnamefont
  {Tveten}}, \bibinfo {author} {\bibfnamefont {A.}~\bibnamefont {Qaiumzadeh}},
  \ and\ \bibinfo {author} {\bibfnamefont {A.}~\bibnamefont {Brataas}},\
  }\href@noop {} {\bibfield  {journal} {\bibinfo  {journal} {Phys. Rev. Lett.}\
  }\textbf {\bibinfo {volume} {112}},\ \bibinfo {pages} {147204} (\bibinfo
  {year} {2014})}\BibitemShut {NoStop}%
\bibitem [{\citenamefont {Kim}\ \emph {et~al.}(2014)\citenamefont {Kim},
  \citenamefont {Tserkovnyak},\ and\ \citenamefont
  {Tchernyshyov}}]{kim_propulsion_2014}%
  \BibitemOpen
  \bibfield  {author} {\bibinfo {author} {\bibfnamefont {S.~K.}\ \bibnamefont
  {Kim}}, \bibinfo {author} {\bibfnamefont {Y.}~\bibnamefont {Tserkovnyak}}, \
  and\ \bibinfo {author} {\bibfnamefont {O.}~\bibnamefont {Tchernyshyov}},\
  }\href {\doibase 10.1103/PhysRevB.90.104406} {\bibfield  {journal} {\bibinfo
  {journal} {Phys. Rev. B}\ }\textbf {\bibinfo {volume} {90}},\ \bibinfo
  {pages} {104406} (\bibinfo {year} {2014})}\BibitemShut {NoStop}%
\bibitem [{\citenamefont {Wang}\ \emph {et~al.}(2015)\citenamefont {Wang},
  \citenamefont {Albert}, \citenamefont {Beg}, \citenamefont {Bisotti},
  \citenamefont {Chernyshenko}, \citenamefont {Cortes-Ortuno}, \citenamefont
  {Hawke},\ and\ \citenamefont {Fangohr}}]{wang_magnon-driven_2015}%
  \BibitemOpen
  \bibfield  {author} {\bibinfo {author} {\bibfnamefont {W.}~\bibnamefont
  {Wang}}, \bibinfo {author} {\bibfnamefont {M.}~\bibnamefont {Albert}},
  \bibinfo {author} {\bibfnamefont {M.}~\bibnamefont {Beg}}, \bibinfo {author}
  {\bibfnamefont {M.-A.}\ \bibnamefont {Bisotti}}, \bibinfo {author}
  {\bibfnamefont {D.}~\bibnamefont {Chernyshenko}}, \bibinfo {author}
  {\bibfnamefont {D.}~\bibnamefont {Cortes-Ortuno}}, \bibinfo {author}
  {\bibfnamefont {I.}~\bibnamefont {Hawke}}, \ and\ \bibinfo {author}
  {\bibfnamefont {H.}~\bibnamefont {Fangohr}},\ }\href@noop {} {\bibfield
  {journal} {\bibinfo  {journal} {Phys. Rev. Lett.}\ }\textbf {\bibinfo
  {volume} {114}},\ \bibinfo {pages} {087203} (\bibinfo {year}
  {2015})}\BibitemShut {NoStop}%
\bibitem [{\citenamefont {Qaiumzadeh}\ \emph {et~al.}(2017)\citenamefont
  {Qaiumzadeh}, \citenamefont {Kristiansen},\ and\ \citenamefont
  {Brataas}}]{qaiumzadeh_controlling_2017}%
  \BibitemOpen
  \bibfield  {author} {\bibinfo {author} {\bibfnamefont {A.}~\bibnamefont
  {Qaiumzadeh}}, \bibinfo {author} {\bibfnamefont {L.~A.}\ \bibnamefont
  {Kristiansen}}, \ and\ \bibinfo {author} {\bibfnamefont {A.}~\bibnamefont
  {Brataas}},\ }\href {http://arxiv.org/abs/1705.01572} {\bibfield  {journal}
  {\bibinfo  {journal} {arXiv:1705.01572 [cond-mat]}\ } (\bibinfo {year}
  {2017})}\BibitemShut {NoStop}%
\bibitem [{\citenamefont {Yang}\ \emph {et~al.}(2015)\citenamefont {Yang},
  \citenamefont {Ryu},\ and\ \citenamefont {Parkin}}]{yang_domain_2015}%
  \BibitemOpen
  \bibfield  {author} {\bibinfo {author} {\bibfnamefont {S.-H.}\ \bibnamefont
  {Yang}}, \bibinfo {author} {\bibfnamefont {K.-S.}\ \bibnamefont {Ryu}}, \
  and\ \bibinfo {author} {\bibfnamefont {S.}~\bibnamefont {Parkin}},\
  }\href@noop {} {\bibfield  {journal} {\bibinfo  {journal} {Nat. Nano.}\
  }\textbf {\bibinfo {volume} {10}},\ \bibinfo {pages} {221} (\bibinfo {year}
  {2015})}\BibitemShut {NoStop}%
\bibitem [{\citenamefont {Gomonay}\ \emph {et~al.}(2016)\citenamefont
  {Gomonay}, \citenamefont {Jungwirth},\ and\ \citenamefont
  {Sinova}}]{gomonay_high_2016}%
  \BibitemOpen
  \bibfield  {author} {\bibinfo {author} {\bibfnamefont {O.}~\bibnamefont
  {Gomonay}}, \bibinfo {author} {\bibfnamefont {T.}~\bibnamefont {Jungwirth}},
  \ and\ \bibinfo {author} {\bibfnamefont {J.}~\bibnamefont {Sinova}},\
  }\href@noop {} {\bibfield  {journal} {\bibinfo  {journal} {Phys. Rev. Lett.}\
  }\textbf {\bibinfo {volume} {117}},\ \bibinfo {pages} {017202} (\bibinfo
  {year} {2016})}\BibitemShut {NoStop}%
\bibitem [{\citenamefont {Shiino}\ \emph {et~al.}(2016)\citenamefont {Shiino},
  \citenamefont {Oh}, \citenamefont {Haney}, \citenamefont {Lee}, \citenamefont
  {Go}, \citenamefont {Park},\ and\ \citenamefont
  {Lee}}]{shiino_antiferromagnetic_2016}%
  \BibitemOpen
  \bibfield  {author} {\bibinfo {author} {\bibfnamefont {T.}~\bibnamefont
  {Shiino}}, \bibinfo {author} {\bibfnamefont {S.-H.}\ \bibnamefont {Oh}},
  \bibinfo {author} {\bibfnamefont {P.~M.}\ \bibnamefont {Haney}}, \bibinfo
  {author} {\bibfnamefont {S.-W.}\ \bibnamefont {Lee}}, \bibinfo {author}
  {\bibfnamefont {G.}~\bibnamefont {Go}}, \bibinfo {author} {\bibfnamefont
  {B.-G.}\ \bibnamefont {Park}}, \ and\ \bibinfo {author} {\bibfnamefont
  {K.-J.}\ \bibnamefont {Lee}},\ }\href@noop {} {\bibfield  {journal} {\bibinfo
   {journal} {Phys. Rev. Lett.}\ }\textbf {\bibinfo {volume} {117}},\ \bibinfo
  {pages} {087203} (\bibinfo {year} {2016})}\BibitemShut {NoStop}%
\bibitem [{\citenamefont {Lan}\ \emph {et~al.}(2015)\citenamefont {Lan},
  \citenamefont {Yu}, \citenamefont {Wu},\ and\ \citenamefont
  {Xiao}}]{lan_spin-wave_2015}%
  \BibitemOpen
  \bibfield  {author} {\bibinfo {author} {\bibfnamefont {J.}~\bibnamefont
  {Lan}}, \bibinfo {author} {\bibfnamefont {W.}~\bibnamefont {Yu}}, \bibinfo
  {author} {\bibfnamefont {R.}~\bibnamefont {Wu}}, \ and\ \bibinfo {author}
  {\bibfnamefont {J.}~\bibnamefont {Xiao}},\ }\href@noop {} {\bibfield
  {journal} {\bibinfo  {journal} {Phys. Rev. X}\ }\textbf {\bibinfo {volume}
  {5}},\ \bibinfo {pages} {041049} (\bibinfo {year} {2015})}\BibitemShut
  {NoStop}%
\bibitem [{\citenamefont {Yu}\ \emph {et~al.}(tion)\citenamefont {Yu},
  \citenamefont {Lan},\ and\ \citenamefont {Xiao}}]{yu_logic_2017}%
  \BibitemOpen
  \bibfield  {author} {\bibinfo {author} {\bibfnamefont {W.}~\bibnamefont
  {Yu}}, \bibinfo {author} {\bibfnamefont {J.}~\bibnamefont {Lan}}, \ and\
  \bibinfo {author} {\bibfnamefont {J.}~\bibnamefont {Xiao}},\ }\href@noop {}
  {\  (\bibinfo {year} {in preparation})}\BibitemShut {NoStop}%
\bibitem [{\citenamefont {Cheng}\ \emph {et~al.}(2014)\citenamefont {Cheng},
  \citenamefont {Xiao}, \citenamefont {Niu},\ and\ \citenamefont
  {Brataas}}]{cheng_spin_2014}%
  \BibitemOpen
  \bibfield  {author} {\bibinfo {author} {\bibfnamefont {R.}~\bibnamefont
  {Cheng}}, \bibinfo {author} {\bibfnamefont {J.}~\bibnamefont {Xiao}},
  \bibinfo {author} {\bibfnamefont {Q.}~\bibnamefont {Niu}}, \ and\ \bibinfo
  {author} {\bibfnamefont {A.}~\bibnamefont {Brataas}},\ }\href {\doibase
  10.1103/PhysRevLett.113.057601} {\bibfield  {journal} {\bibinfo  {journal}
  {Phys. Rev. Lett.}\ }\textbf {\bibinfo {volume} {113}},\ \bibinfo {pages}
  {057601} (\bibinfo {year} {2014})}\BibitemShut {NoStop}%
\bibitem [{\citenamefont {Cheng}\ \emph {et~al.}(2016)\citenamefont {Cheng},
  \citenamefont {Daniels}, \citenamefont {Zhu},\ and\ \citenamefont
  {Xiao}}]{cheng_antiferromagnetic_2016}%
  \BibitemOpen
  \bibfield  {author} {\bibinfo {author} {\bibfnamefont {R.}~\bibnamefont
  {Cheng}}, \bibinfo {author} {\bibfnamefont {M.~W.}\ \bibnamefont {Daniels}},
  \bibinfo {author} {\bibfnamefont {J.-G.}\ \bibnamefont {Zhu}}, \ and\
  \bibinfo {author} {\bibfnamefont {D.}~\bibnamefont {Xiao}},\ }\href@noop {}
  {\bibfield  {journal} {\bibinfo  {journal} {Sci. Rep.}\ }\textbf {\bibinfo
  {volume} {6}},\ \bibinfo {pages} {24223} (\bibinfo {year}
  {2016})}\BibitemShut {NoStop}%
\bibitem [{\citenamefont {Lan}\ \emph {et~al.}(2017)\citenamefont {Lan},
  \citenamefont {Yu},\ and\ \citenamefont {Xiao}}]{lan_antiferromagnetic_2017}%
  \BibitemOpen
  \bibfield  {author} {\bibinfo {author} {\bibfnamefont {J.}~\bibnamefont
  {Lan}}, \bibinfo {author} {\bibfnamefont {W.}~\bibnamefont {Yu}}, \ and\
  \bibinfo {author} {\bibfnamefont {J.}~\bibnamefont {Xiao}},\ }\href {\doibase
  10.1038/s41467-017-00265-5} {\bibfield  {journal} {\bibinfo  {journal} {Nat.
  Commun.}\ }\textbf {\bibinfo {volume} {8}},\ \bibinfo {pages} {178} (\bibinfo
  {year} {2017})}\BibitemShut {NoStop}%
\bibitem [{\citenamefont {Goldstein}(2003)}]{Goldstein_2003_polarized}%
  \BibitemOpen
  \bibfield  {author} {\bibinfo {author} {\bibfnamefont {D.}~\bibnamefont
  {Goldstein}},\ }\href@noop {} {\emph {\bibinfo {title} {Polarized Light
  Second Edition, Revised and Expanded}}}\ (\bibinfo  {publisher} {Marcel
  Dekker, New York},\ \bibinfo {year} {2003})\BibitemShut {NoStop}%
\bibitem [{\citenamefont {Tveten}\ \emph {et~al.}(2013)\citenamefont {Tveten},
  \citenamefont {Qaiumzadeh}, \citenamefont {Tretiakov},\ and\ \citenamefont
  {Brataas}}]{tveten_staggered_2013}%
  \BibitemOpen
  \bibfield  {author} {\bibinfo {author} {\bibfnamefont {E.~G.}\ \bibnamefont
  {Tveten}}, \bibinfo {author} {\bibfnamefont {A.}~\bibnamefont {Qaiumzadeh}},
  \bibinfo {author} {\bibfnamefont {O.~A.}\ \bibnamefont {Tretiakov}}, \ and\
  \bibinfo {author} {\bibfnamefont {A.}~\bibnamefont {Brataas}},\ }\href@noop
  {} {\bibfield  {journal} {\bibinfo  {journal} {Phys. Rev. Lett.}\ }\textbf
  {\bibinfo {volume} {110}},\ \bibinfo {pages} {127208} (\bibinfo {year}
  {2013})}\BibitemShut {NoStop}%
\bibitem [{\citenamefont {Tveten}\ \emph {et~al.}(2016)\citenamefont {Tveten},
  \citenamefont {M\"{u}ller}, \citenamefont {Linder},\ and\ \citenamefont
  {Brataas}}]{tveten_intrinsic_2016}%
  \BibitemOpen
  \bibfield  {author} {\bibinfo {author} {\bibfnamefont {E.~G.}\ \bibnamefont
  {Tveten}}, \bibinfo {author} {\bibfnamefont {T.}~\bibnamefont {M\"{u}ller}},
  \bibinfo {author} {\bibfnamefont {J.}~\bibnamefont {Linder}}, \ and\ \bibinfo
  {author} {\bibfnamefont {A.}~\bibnamefont {Brataas}},\ }\href {\doibase
  10.1103/PhysRevB.93.104408} {\bibfield  {journal} {\bibinfo  {journal} {Phys.
  Rev. B}\ }\textbf {\bibinfo {volume} {93}},\ \bibinfo {pages} {104408}
  (\bibinfo {year} {2016})}\BibitemShut {NoStop}%
\bibitem [{\citenamefont {Kittel}(1951)}]{kittel_theory_1951}%
  \BibitemOpen
  \bibfield  {author} {\bibinfo {author} {\bibfnamefont {C.}~\bibnamefont
  {Kittel}},\ }\href@noop {} {\bibfield  {journal} {\bibinfo  {journal} {Phys.
  Rev.}\ }\textbf {\bibinfo {volume} {82}},\ \bibinfo {pages} {565} (\bibinfo
  {year} {1951})}\BibitemShut {NoStop}%
\bibitem [{\citenamefont {Keffer}\ and\ \citenamefont
  {Kittel}(1952)}]{keffer_theory_1952}%
  \BibitemOpen
  \bibfield  {author} {\bibinfo {author} {\bibfnamefont {F.}~\bibnamefont
  {Keffer}}\ and\ \bibinfo {author} {\bibfnamefont {C.}~\bibnamefont
  {Kittel}},\ }\href@noop {} {\bibfield  {journal} {\bibinfo  {journal} {Phys.
  Rev.}\ }\textbf {\bibinfo {volume} {85}},\ \bibinfo {pages} {329} (\bibinfo
  {year} {1952})}\BibitemShut {NoStop}%
\bibitem [{\citenamefont {Haldane}(1983)}]{haldane_nonlinear_1983}%
  \BibitemOpen
  \bibfield  {author} {\bibinfo {author} {\bibfnamefont {F.~D.~M.}\
  \bibnamefont {Haldane}},\ }\href {\doibase 10.1103/PhysRevLett.50.1153}
  {\bibfield  {journal} {\bibinfo  {journal} {Phys. Rev. Lett.}\ }\textbf
  {\bibinfo {volume} {50}},\ \bibinfo {pages} {1153} (\bibinfo {year}
  {1983})}\BibitemShut {NoStop}%
\bibitem [{\citenamefont {Poschl}\ and\ \citenamefont
  {Teller}(1933)}]{poschl_bemerkungen_1933}%
  \BibitemOpen
  \bibfield  {author} {\bibinfo {author} {\bibfnamefont {G.}~\bibnamefont
  {Poschl}}\ and\ \bibinfo {author} {\bibfnamefont {E.}~\bibnamefont
  {Teller}},\ }\href@noop {} {\bibfield  {journal} {\bibinfo  {journal} {Z.
  Phys.}\ }\textbf {\bibinfo {volume} {83}},\ \bibinfo {pages} {143} (\bibinfo
  {year} {1933})}\BibitemShut {NoStop}%
\bibitem [{COM()}]{COMSOL}%
  \BibitemOpen
  \href@noop {} {\enquote {\bibinfo {title} {\text{COMSOL Multiphysics}},}\
  }\bibinfo {howpublished} {\url{http://comsol.com/}}\BibitemShut {NoStop}%
\end{thebibliography}

%

 \end{document}